\documentclass[showpacs,twocolumn,prb]{revtex4-2}
\usepackage{amsfonts}
\usepackage{graphicx}
\usepackage{amsmath}
\usepackage{amssymb}

\begin{document}

\title{Longitudinal and spin/valley Hall optical conductivity in single
layer $MoS_{2}$}
\author{Zhou Li$^1$}
\email{lizhou@univmail.cis.mcmaster.ca}
\author{J. P. Carbotte$^{1,2}$}
\email{carbotte@univmail.cis.mcmaster.ca}

\begin{abstract}
A monolayer of $MoS_{2}$ has a non-centrosymmetric crystal structure, with
spin polarized bands. It is a two valley semiconductor with direct gap
falling in the visible range of the electromagnetic spectrum. Its optical
properties are of particular interest in relation to valleytronics and
possible device applications. We study the longitudinal and the transverse
Hall dynamical conductivity which is decomposed into charge, spin and valley
contributions. Circular polarized light associated with each of the two
valleys separately is considered and results are filtered according to spin
polarization. Temperature can greatly change the spin admixture seen in the
frequency window where they are not closely in balance.
\end{abstract}

\date{\today }
\maketitle

\affiliation{$^1$
Department of Physics, McMaster University, Hamilton, Ontario,
Canada,L8S 4M1 \\
$^2$ Canadian Institute for Advanced Research, Toronto, Ontario,
Canada M5G 1Z8}

\section{Introduction}

Since the isolation of graphene,\cite{Novo1,Zhang} the search for new two
dimensional atomic membranes possibly with novel functionalities has
intensified. $MoS_{2}$\cite{Novo2} in its single layered form is a two
valley direct band gap semiconductor with gap in the visible and so is of
interest for device applications. Through excitation by right and left hand
polarized light excess populations of a selected valley can be generated
which make this material ideal for valleytronics.\cite{Xiao,Zeng,Mak,Cao1}
Because the two inequivalent valleys are separated in the Brillouin zone by
a large momentum, intervalley scattering should be small. Consequently the
valley index becomes a new degree of freedom analogous to spin in
semiconductors. Just like manipulating spin has lead to spintronics,\cite%
{Wolf,Fabian} manipulating valley index can produce new effects including
using it to carry information. As an example, in the context of graphene
Xiao et.al.\cite{Xiao1} showed that a contrasting intrinsic magnetic moment
and Berry curvature can be associated with carrier valley index. As a second
example a valley filter device is described by Rycerz et.al.\cite{Rycerz}

An important issue is possible pathways to achieve valley polarization i.e.
populating states preferentially in one valley.\cite{Behnia} One way
discussed theoretically\cite{Xiao1,Xiao,Yao,Feng} is to use circular
polarized light and this has been demonstrated recently by three
experimental groups.\cite{Behnia,Zeng,Mak,Cao1} Selection rules on the
absorption of right (or left) handed polarized light assures that this
radiation excites almost exclusively charge carriers residing in a single
valley with index $\tau=+1$ (or $-1$). The physical quantity that comes into
the description of such processes is the real part of the AC optical
conductivity for right (+1) or left (-1) polarization $Re\sigma_{\pm }$ as a
function of photon energy $\omega$. In terms of longitudinal $%
\sigma_{xx}(\omega)$ and transverse $\sigma_{xy}(\omega)$ conductivity $%
\sigma_{\pm }(\omega )\equiv\sigma _{xx}(\omega )\pm i\sigma _{xy}(\omega )$%
. The conductivity also plays an important role in determining optical
properties of nanostructures as we can see from the references [%
\onlinecite{Jablan},\onlinecite{Nikitin},\onlinecite{Vakil},%
\onlinecite{Koppens},\onlinecite{Chen},\onlinecite{Fei}]. In order to solve
the Maxwell equations for systems with graphene sheet between two media with
different dielectric constants, one has to know the conductivity of two
dimensional graphene. Results for the conductivity of single layer $MoS_{2}$
will be useful if similar devices were built from $MoS_{2}$ instead of
graphene.

$MoS_{2}$ is a layer of molybdenum atoms between two layers of sulfur in a
trigonal prismatic arrangement which does not have inversion symmetry. In
momentum space at the $K$ and $-K$ points of the honeycomb lattice\cite%
{Zhu,Li} the valence and conduction bands are separated by a large
semiconductor direct band gap $\Delta $ \cite{Mak1} and there is a large
spin orbit coupling leading to a spin polarization of the valence band, spin
up $\uparrow $ and down $\downarrow $ as the $z$ -component of the spin
operator $s_{z}$ commutes with the Hamiltonian and hence remains a good
quantum number. A minimal Hamiltonian which describes the band structure of $%
MoS_{2}$ (valid near the main absorption edge)is found in reference \cite%
{Xiao} with parameters based on first principle calculations for the
group-VI dichalcogenides.\cite{Zhu,Chei} A more complete theory is found in
the reference [\onlinecite{Feng}] where a calculation of the dc Hall
conductivity and Berry curvature over the entire Brillouin zone is
presented. Part of the Hamiltonian describes the dynamics of massive Dirac
fermions which are known from the graphene literature to have an optical
response quite different from that of an ordinary 2D electron gas. For
example a universal constant background conductivity of $e^{2}/(4\hbar)$
\cite{Castro,Kotov}(for a single spin and single valley this constant
becomes $e^{2}/(16\hbar)$) is predicted and observed for photon energy $%
\omega$ greater than twice the chemical potential $\mu$ for massless Dirac
fermions. We use this Hamiltonian to calculate the dynamic optical
conductivity as a function of photon energy to several electron volts.
Results are presented for longitudinal $\sigma _{xx}(\omega ) $ as well as
transverse conductivity $\sigma _{xy}(\omega )$ which is separated in
charge, spin and valley Hall conductivity. Appropriate contributions from
different spin channel are presented separately and special attention is
payed to the effects of right and left handed light polarization.
Temperature effects are also considered. In section II we present the
Hamiltonian and the Green's function on which our calculations are based. In
section III the mathematical expressions for the conductivity are evaluated
and results for spin and valley Hall conductivity are given for several
representative values of the chemical potential. Results for circular
polarized light are found in section IV and section V contains a summary.

\section{Formalism}

The Hamiltonian for $MoS_{2}$ at $K$ and $-K$ points is

\begin{equation}
H_{0}=at(\tau k_{x}\hat{\sigma}_{x}+k_{y}\hat{\sigma}_{y})+\frac{\Delta }{2}
\hat{\sigma}_{z}-\lambda \tau \frac{\hat{\sigma}_{z}-1}{2}\hat{S}_{z}
\label{Hamiltonian}
\end{equation}
with $2\lambda $ the spin orbit splitting at the top of the valence band and
we take $2\lambda =0.15eV$, $a$ the lattice parameter $3.193\mathring{A}$, $t
$ the hopping $t=1.1eV$, $\hat{\sigma}$ the Pauli matrices and $\hat{S}_{z} $
the spin matrix for the z-component of spin $s_{z}$ which is a good quantum
number here. The index $\tau =\pm 1$ is the valley $K(-K)$ respectively and $%
\Delta $ is the direct band gap equal to $1.66eV$ between valence and
conduction bands. The $x$ and $y$ velocity components based on (\ref%
{Hamiltonian}) are
\begin{equation}
v_{x}=\frac{\partial H_{0}}{\hbar \partial k_{x}}=\frac{at}{\hbar }\tau
\sigma _{x},v_{y}=\frac{\partial H_{0}}{\hbar \partial k_{y}}=\frac{at}{
\hbar }\sigma _{y}.
\end{equation}

Here we will be interested in charge, spin and valley current given
respectively as $\mathbf{j}=e\mathbf{v,j}^{s}=\frac{\hbar }{2}s_{z}\mathbf{v}
$ and $\mathbf{j}^{v}=\tau \mathbf{v}$. The eigen energies and vectors of ( %
\ref{Hamiltonian}) are
\begin{equation}
E_{k}^{\pm }(\tau ,s_{z})=\lambda \tau s_{z}/2\pm \sqrt{a^{2}t^{2}k^{2}+(%
\frac{\Delta ^{\prime }}{2})^{2}}  \label{Ek}
\end{equation}%
and
\begin{equation}
u_{n}(k)=\frac{atk\left( 1,\frac{-\Delta ^{\prime }/2\pm \sqrt{%
a^{2}t^{2}k^{2}+(\Delta ^{\prime }/2)^{2}}}{at}\frac{\tau k_{x}+ik_{y}}{k^{2}%
}\right) ^{T}}{\sqrt{a^{2}t^{2}k^{2}+(-\Delta ^{\prime }/2\pm \sqrt{%
a^{2}t^{2}k^{2}+(\Delta ^{\prime }/2)^{2}})^{2}}}
\end{equation}%
with $\Delta ^{\prime }=\Delta -\lambda \tau s_{z}$.\cite{Nicol} For later
reference the Berry curvature for the conduction band $E_{k}^{+}$ is
\begin{eqnarray}
&&\Omega _{c}(k)=z\cdot \nabla _{k}\times \left\langle u_{n}(k)|i\nabla
_{k}|u_{n}(k)\right\rangle   \notag \\
&=&-\tau \frac{2a^{2}t^{2}\Delta ^{\prime }}{[\Delta ^{\prime
2}+4a^{2}t^{2}k^{2}]^{3/2}}
\end{eqnarray}%
with valence band $\Omega _{v}(k)=-\Omega _{c}(k)$. With these solutions the
green's function $\hat{G}_{0}(\mathbf{k},i\omega _{n})$ with Matsubara
frequencies $\omega _{n}$ is
\begin{widetext}
\begin{equation}
\hat{G}_{0}(\mathbf{k},i\omega _{n})=\frac{(i\hbar \omega _{n}+\mu
-\lambda \tau s_{z}/2)\hat{I}+(\frac{\Delta ^{\prime
}}{2})\hat{\sigma}_{z}+at(\tau k_{x}\sigma _{x}+k_{y}\sigma _{y})}{
(i\hbar \omega _{n}+\mu -\lambda \tau
s_{z}/2)^{2}-a^{2}t^{2}k^{2}-(\frac{ \Delta ^{\prime
}}{2})^{2}}\equiv G_{I}(\mathbf{k},i\omega
_{n})\hat{I}+G_{z}\hat{\sigma}_{z}+G_{x}
\hat{\sigma}_{x}+G_{y}\hat{\sigma}_{y}
\end{equation}
\end{widetext}which gives
\begin{eqnarray*}
&&G_{I}(\mathbf{k},i\omega _{n}) \\
&=&\frac{1}{2}\frac{1}{i\hbar \omega _{n}+\mu -\lambda \tau s_{z}/2-\sqrt{%
a^{2}t^{2}k^{2}+(\frac{\Delta ^{\prime }}{2})^{2}}} \\
&&+\frac{1}{2}\frac{1}{i\hbar \omega _{n}+\mu -\lambda \tau s_{z}/2+\sqrt{%
a^{2}t^{2}k^{2}+(\frac{\Delta ^{\prime }}{2})^{2}}}
\end{eqnarray*}

\begin{eqnarray}
G_{z} &=&\frac{\frac{\Delta ^{\prime }}{2}}{(i\hbar \omega _{n}+\mu -\lambda
\tau s_{z}/2)}G_{I}(\mathbf{k},i\omega _{n})  \notag \\
G_{x} &=&\frac{at\tau k_{x}}{(i\hbar \omega _{n}+\mu -\lambda \tau s_{z}/2)}
G_{I}(\mathbf{k},i\omega _{n})  \notag \\
G_{y} &=&\frac{atk_{y}}{(i\hbar \omega _{n}+\mu -\lambda \tau s_{z}/2)}%
G_{I}( \mathbf{k},i\omega _{n}).
\end{eqnarray}

The density of state $N(\epsilon )$ is given by
\begin{equation}
N(\epsilon )=-\frac{1}{\pi }\sum_{\mathbf{k}}ImG_{I}(\mathbf{k},i\omega
_{n}->\epsilon +i\delta ).
\end{equation}
where $\sum $ is a sum over momentum, $Im$ means taking the imaginary part.
The longitudinal conductivity $\sigma _{xx}(\omega )$, charge Hall
conductivity $\sigma _{xy}(\omega )$, spin Hall conductivity $\sigma
_{xy}^{s}(\omega )$ and valley Hall conductivity $\sigma _{xy}^{v}(\omega )$
are given by\cite{Carbotte1,Carbotte2,Peres}

\begin{eqnarray*}
&&\sigma _{xx}(\omega )=-\frac{e^{2}a^{2}t^{2}}{i\omega \hbar ^{2}}T\sum_{
\mathbf{k,}l} \\
&&Tr\langle \sigma _{x}G_{0}(\mathbf{k,}i\omega _{l})\sigma _{x}G_{0}(
\mathbf{k,}i\omega _{l}+i\omega _{n})\rangle _{i\omega _{n}->\omega +i\delta
}
\end{eqnarray*}
\begin{eqnarray*}
&&\sigma _{xy}(\omega )=-\frac{e^{2}a^{2}t^{2}}{i\omega \hbar ^{2}}T\sum_{
\mathbf{k,}l} \\
&&Tr\langle \tau \sigma _{x}G_{0}(\mathbf{k,}i\omega _{l}) \sigma _{y}G_{0}(
\mathbf{k,}i\omega _{l}+i\omega _{n})\rangle _{i\omega _{n}->\omega +i\delta
}
\end{eqnarray*}
\begin{eqnarray*}
&&\sigma _{xy}^{s}(\omega )=-\frac{ea^{2}t^{2}}{i\omega \hbar ^{2}}T\sum_{
\mathbf{k,}l} \\
&&Tr\langle \frac{\hbar }{2}s_{z}\tau \sigma _{x}G_{0}(\mathbf{k,}i\omega
_{l})\sigma _{y}G_{0}(\mathbf{k,}i\omega _{l}+i\omega _{n})\rangle _{i\omega
_{n}->\omega +i\delta }
\end{eqnarray*}
\begin{eqnarray}
&&\sigma _{xy}^{v}(\omega )=-\frac{ea^{2}t^{2}}{i\omega \hbar ^{2}}T\sum_{
\mathbf{k,}l}  \notag \\
&&Tr\langle \sigma _{x}G_{0}(\mathbf{k,}i\omega _{l}) \sigma _{y}G_{0}(%
\mathbf{k,}i\omega _{l}+i\omega _{n})\rangle _{i\omega _{n}->\omega +i\delta
}
\end{eqnarray}

After simplification we get
\begin{subequations}
\begin{equation}
\sigma _{xy}(\omega )=-\frac{e^{2}}{\hbar}\sum_{\mathbf{k,}\tau
,s_{z}}\Omega _{c}(k)[f(E^{+})-f(E^{-})]g  \label{Hall}
\end{equation}
with
\begin{equation}
g\equiv \frac{(4a^{2}t^{2}k^{2}+\Delta ^{\prime }{}^{2})}{(\hbar \omega
+i\delta )^{2}-(4a^{2}t^{2}k^{2}+\Delta ^{\prime 2})}  \label{g}
\end{equation}
for spin and valley Hall conductivity, $g$ in (\ref{g}) is to be replaced by
$s_{z}g/e$ and $\tau g/e$ respectively. Here $f(x)$ is the Fermi Dirac
function, which contains the chemical potential $\mu$. The longitudinal
conductivity is

\begin{eqnarray}
&&\sigma _{xx}(\omega )=\frac{e^{2}}{\hbar ^{2}}\sum_{\mathbf{k,}\tau
,s_{z}} \frac{2a^{2}t^{2}(4a^{2}t^{2}k_{y}^{2}+\Delta ^{\prime }{}^{2})}{%
i\omega \lbrack \Delta ^{\prime 2}+4a^{2}t^{2}k^{2}]^{1/2}}  \notag \\
&&\times \frac{\lbrack f(E^{+})-f(E^{-})]}{(\hbar \omega +i\delta
)^{2}-(4a^{2}t^{2}k^{2}+\Delta ^{\prime 2})}  \label{xx}
\end{eqnarray}

To obtain (\ref{xx}) we have included only the interband transitions. There
is an additional intraband contribution which provides a delta function
contribution at $\omega=0$. When residual scattering is included in the
calculations, the intraband piece broadens into a Drude peak which can
overlap with the interband contribution. But in the pure limit which is the
case we are considering here we need not consider this contribution. What
replaces the Berry curvature in the expression for the longitudinal
conductivity $\sigma _{xx}$ is a factor

\begin{equation}
h\equiv -\frac{2a^{2}t^{2}(4a^{2}t^{2}k_{y}^{2}+\Delta ^{\prime }{}^{2})}{
i\omega \lbrack \Delta ^{\prime 2}+4a^{2}t^{2}k^{2}]^{3/2}}
\end{equation}
For $k=0$ this factor in $\sigma _{xx}$ and $\sigma _{xy}$ agree and as we
will see later this leads to a valley selection rule for circular polarized
light; at finite $k$ however the cancelation is no longer exact.

\section{Results for spin and valley Hall conductivity}

\begin{figure}[tp]
\begin{center}
\includegraphics[height=8in,width=3.0in]{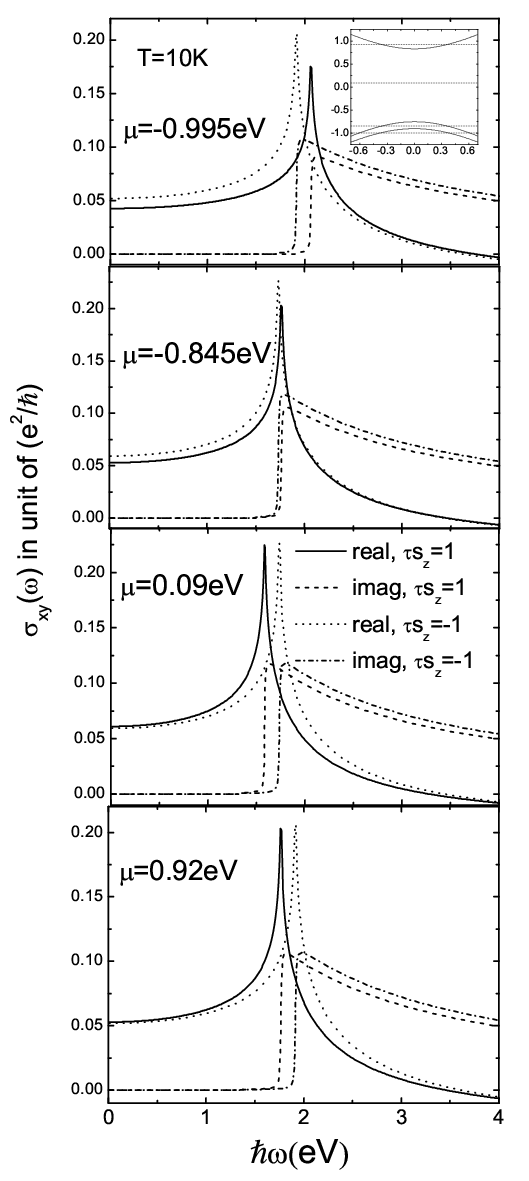}
\end{center}
\caption{Real (solid) part of the Hall conductivity $\protect\sigma _{xy}$
vs. $\protect\omega$, and its imaginary part (dashed) for $\protect\tau %
s_{z}=1$. The dotted and dashed dotted curves are for $\protect\tau s_{z}=-1$
instead. There are 4 frames from top to bottom for four values of chemical
potential $\protect\mu$ namely $-0.995, -0.845, 0.09$ and $0.92 eV$
respectively. The inset provides a sketch of the bands in $MoS_{2}$ and how
the four values of $\protect\mu$ relate to these. In all cases the bands are
cut off at momentum $ka=3.0$.}
\label{fig1}
\end{figure}

\begin{figure}[tp]
\begin{center}
\includegraphics[height=8.0in,width=3.0in]{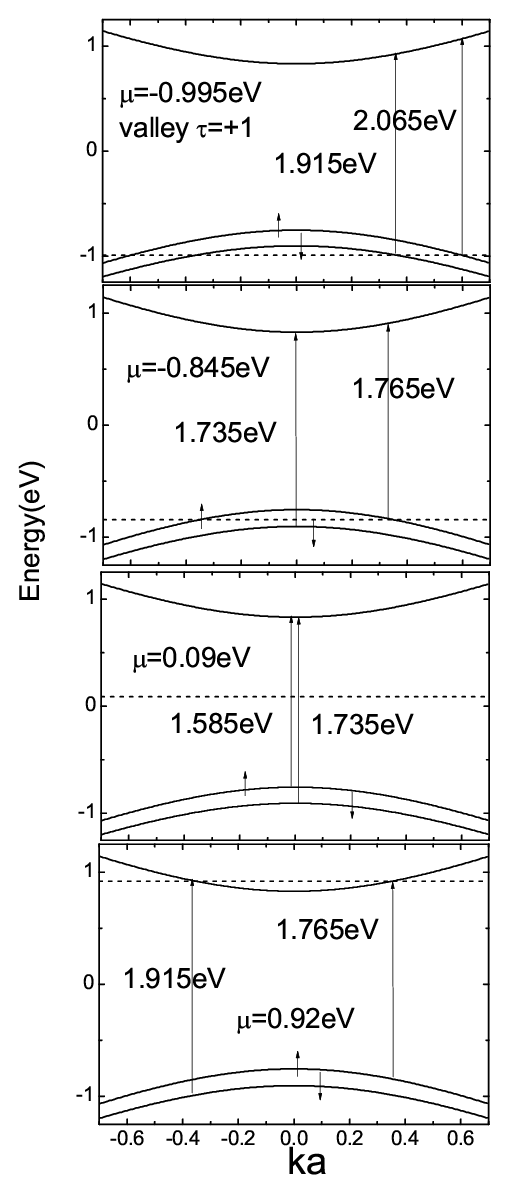}
\end{center}
\caption{The energy bands (solid) for $MoS_{2}$ with chemical potential $%
\protect\mu$ level indicated by a dashed horizontal line. The extremal
optical transitions are also indicated in each case. From top to bottom
frame $\protect\mu=-0.995,-0.845,0.09$ and $0.92 eV$.}
\label{fig2}
\end{figure}

In Figure \ref{fig1} we show our results for the Hall conductivity $\sigma
_{xy}(\omega )$ vs. $\hbar \omega $ in units of $e/\hbar $ for the four
values of the chemical potential shown in the inset of the top frame where
the band energies $E_{k}^{\pm }(\tau ,s_{z})$ are sketched (equation (\ref%
{Ek})). $\mu =-0.995eV$ falls below the top of the lowest energy spin
polarized valence band; $\mu =-0.845eV$ falls between the two valence band; $%
\mu =0.09eV$ is between valence and conduction band(insulator) while $\mu
=0.92eV$ falls in the conduction band also spin split but this splitting is
small. In the main frame which has 4 frames each for a different value of $%
\mu $ we show separately the contribution for $\tau s_{z}=\pm 1$ with $%
\sigma _{xy}(\omega )$ the sum over both index $s_{z}$(spin) and $\tau $
(valley). The solid line is the real part for $\tau s_{z}=1$ while for $\tau
s_{z}=-1$ the dotted line applies. The dashed and dash-dotted are for their
respective imaginary part. We note a sharp onset in these two last curves
the energy of which can be traced to the minimum energy associated with
possible interband optical transition as shown in Figure \ref{fig2}. As an
example, in the top frame we see that, for $\mu =-0.995eV$, the onset of the
interband transitions for spin $\uparrow $ occurs at higher energies than
for spin $\downarrow $ (here the valley index has been taken to be $1$ ).
Corresponding to the onset in $Im\sigma _{xy}$ there is a peak in its real
part at this same energy, as they are related by the Kramers-Kronig
relations. The results presented were obtained through numerical evaluation
of equation (\ref{Hall}). The numerical values do have some dependence on
cut off used on the energy $(atk)^{2}$. Here we have set the cut off $ak$ to
be $3$ and restrict ourselves to photon energies below 3eV. In this energy
range choosing a larger cut off makes no difference to the results which
have converged. For photon energies above 3eV, a range not considered
explicitly in our figures, increasing the cut off even to infinity has no
qualitative effect on the results for the imaginary part of $\sigma
_{xy}(\omega )$. It affects the real part more. With a cut off, there is a
zero in $Re\sigma _{xy}(\omega )$ at some high energy. As the cut off is
increased, the energy of the zero in $Re\sigma _{xy}(\omega )$ moves to
higher energies and in the limit of infinite cut off no zero remains for
finite $\omega$ as it has moved to infinity. Taking the infinite band limit
has the advantage that simple analytic expressions can be obtained which can
be useful. For example it is straight forward to show that
\end{subequations}
\begin{eqnarray}
&&Im\sigma _{xy}(\omega )=-\frac{e^{2}}{16\hbar }\frac{2(\Delta -\lambda )}{
\hbar \omega }\theta (\hbar \omega -(\Delta -\lambda ))  \notag \\
&&\times \lbrack f(\frac{\lambda }{2}+\frac{\hbar \omega }{2}-\mu )-f(\frac{
\lambda }{2}-\frac{\hbar \omega }{2}-\mu )]  \label{Imxy}
\end{eqnarray}
where we have made explicit the value of chemical potential attached to the
thermal function $f(x)$ (Fermi Dirac distribution $f(x)\equiv 1/(e^{\beta
x}+1)$ with $\beta =1/T$). Note the onset at $(\Delta -\lambda )$ and the $%
1/\omega $ drop in this function as a function of $\omega $. While there are
some quantitative differences of the form (\ref{Imxy}) with our numerical
results there are no qualitative changes. Another reason for restricting the
range of photon energies considered, as we have done here, is that the model
Hamiltonian (1) is itself valid only near the main absorption edge. A first
principle calculation of the dc Hall conductivity and Berry curvature over
the entire Brillouin zone which goes beyond what we have done is given by
Feng et.al. \cite{Feng} Here we really cannot access accurately the high
energy region.

\begin{figure}[tp]
\begin{center}
\includegraphics[height=8.0in,width=3.0in]{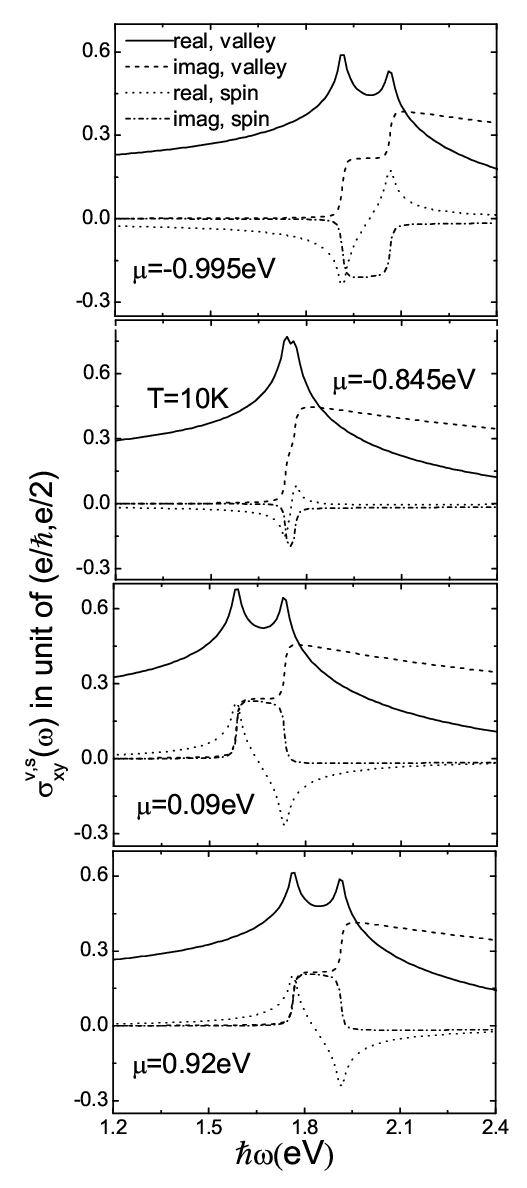}
\end{center}
\caption{The real (solid) and imaginary (dashed) part of the valley Hall
conductivity $\protect\sigma _{xy}$ vs. $\protect\omega$. The dotted and
dashed dotted curves are similar results for the spin Hall conductivity
case. From top to bottom frame the chemical potential $\protect\mu=-0.995,
-0.845, 0.09$ and $0.92 eV$. The inset in Figure \protect\ref{fig1} provides
a sketch of the bands in $MoS_{2}$ and how the four values of $\protect\mu$
relate to these. }
\label{fig3}
\end{figure}

To get the spin and valley Hall conductivity from the results presented in
Figure \ref{fig1} components need to be added according to the weighting $%
\hbar s_{z}/(2e)$ and $\tau /e$ respectively as noted in relation to
equation (\ref{Hall}). Results are presented in Figure \ref{fig3}. The real
part of the valley Hall conductivity is the solid curve with dashed the
corresponding imaginary part, while the dotted and dashed dotted are for the
spin Hall conductivity. The differences in weighting $s_{z}$ and $\tau $ has
a large effect on the resulting Hall conductivity. For example the real part
of the valley Hall conductivity is everywhere positive in our model while
the spin Hall conductivity starts negative at $\omega =0$ and is very small
in comparison to the valley Hall conductivity. Both show peaks at the onset
energies associated with their imaginary part. Because of the spin splitting
of our bands there are two peaks seen clearly in the real part of the valley
Hall conductivity (solid curve), after which it drops off gradually with
increasing $\omega $. Consequently the real part of the spin Hall
conductivity (dotted curve) changes sign with increasing $\omega $, because
the two peaks found in Figure 1 (solid and dotted curves) are associated
with opposite spin, this sign changing behavior has also been found in the
Figure 4(a) of the reference [\onlinecite{Feng}]. While the results show
quantitative variations with value chemical potential (see inset in Figure %
\ref{fig1}), there is no qualitative change.

\begin{figure}[tp]
\begin{center}
\includegraphics[height=4.0in,width=3.0in]{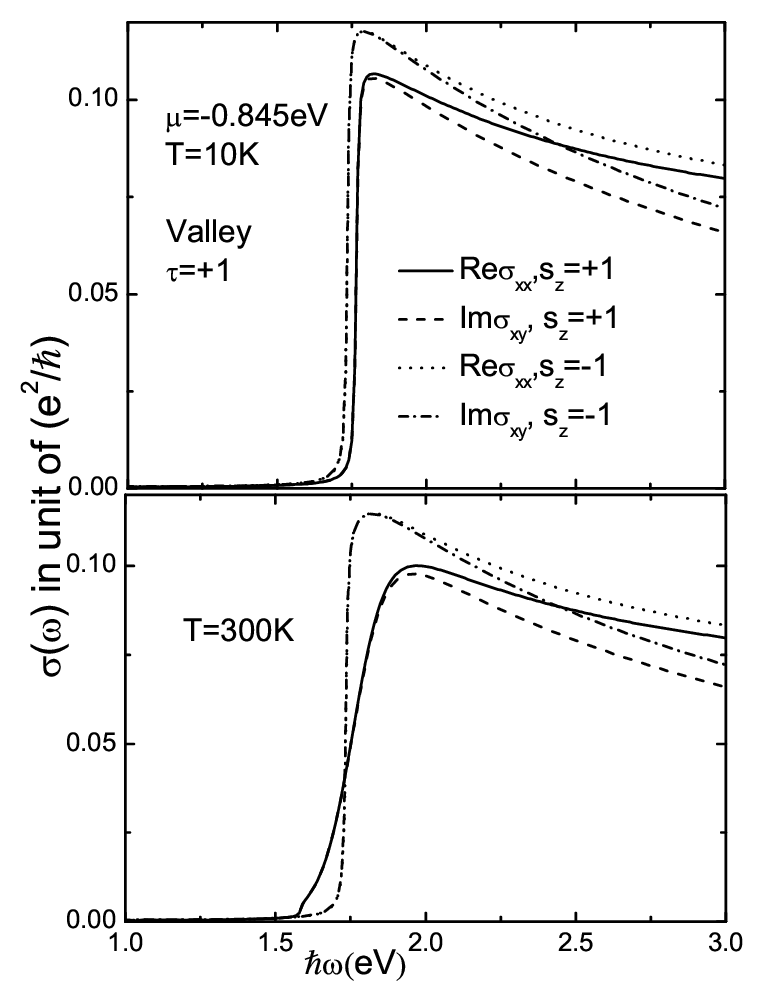}
\end{center}
\caption{The absorptive part of the longitudinal ($Re\protect\sigma_{xx}(
\protect\omega )$) and transverse Hall conductivity ($Im\protect\sigma %
_{xy}( \protect\omega )$) as a function of photon energy $\protect\omega$.
The chemical potential is $\protect\mu=-0.845 eV$. The top frame is for
temperature $T=10K$ and the bottom for $T=300K$. The solid and dashed curves
are for $Re\protect\sigma_{xx}(\protect\omega )$ and $Im\protect\sigma _{xy}(%
\protect\omega )$ with $\protect\tau =+1$ (valley index) and $s_{z}=+1$
(spin index), while dotted and dashed dotted are for $s_{z}=-1 $. }
\label{fig4}
\end{figure}

The zero energy limit ($\omega =0$) of the Hall conductivity is of interest
and can be worked out analytically in the infinite band limit. At zero
temperature ($T=0$) the results are ($\sigma _{xy}^{s}$ in unit of $\frac{e}{
2}$, $\sigma _{xy}^{v}$ in unit of $\frac{e}{\hbar }$)

\begin{subequations}
\begin{eqnarray}
\sigma _{xy}^{s}(\omega &=&0)=\frac{\lambda }{\pi }\frac{\Delta -2\mu }{
\lambda ^{2}-4\mu ^{2}} \\
\sigma _{xy}^{v}(\omega &=&0)=\frac{1}{\pi }\frac{2\mu \Delta -\lambda ^{2}}{
\lambda ^{2}-4\mu ^{2}}
\end{eqnarray}
for $\mu $ in the lowest valence band;
\begin{equation}
\frac{1}{2\pi }\frac{\Delta -2\lambda +2\mu }{\lambda -2\mu }\text{ and }
\frac{1}{2\pi }\frac{\Delta -2\mu }{\lambda -2\mu }
\end{equation}
for $\mu $ between $\uparrow ,\downarrow $ valence band;
\begin{equation}
0\text{ and }\frac{1}{\pi }
\end{equation}
for $\mu $ between valence and conduction band and
\begin{equation}
\frac{\lambda }{\pi }\frac{\Delta -2\mu }{4\mu ^{2}-\lambda ^{2}}\text{ and }
\frac{1}{\pi }\frac{2\mu \Delta -\lambda ^{2}}{4\mu ^{2}-\lambda ^{2}}
\end{equation}
for $\mu $ above the conduction band. Note in particular for $\mu $ between
valence and conduction band $\sigma _{xy}^{s}(\omega =0)=0$ and $\sigma
_{xy}^{v}(\omega =0)=\frac{e}{\pi \hbar }$. Here we do not have a spin Hall
insulating state. Note that for the real part of the Hall conductivity at
zero temperature an analytical expression exists, for example for $\tau
=1,s_{z}=1$,
\end{subequations}
\begin{subequations}
\begin{eqnarray}
&&Re\sigma _{xy}(\omega )=-\frac{e^{2}}{8\pi \hbar ^{2}\omega }(\Delta
-\lambda )  \notag \\
&&\times \ln |\frac{\hbar \omega -2\sqrt{x_{\min }}}{\hbar \omega +2\sqrt{
x_{\min }}}\frac{\hbar \omega +2\sqrt{x_{\max }}}{\hbar \omega -2\sqrt{
x_{\max }}}|
\end{eqnarray}%
with
\begin{eqnarray}
x_{\min } &=&\max [(\mu -\frac{\lambda }{2})^{2},(\frac{\Delta -\lambda }{2}
)^{2}] \\
x_{\max } &=&(ka)_{cut}^{2}t^{2}+(\frac{\Delta -\lambda }{2})^{2}.
\end{eqnarray}
$(ka)_{cut}$ is the cut off for $ka$, in the infinite band approximation $%
(ka)_{cut}\rightarrow \infty $ while in our numerical results $%
(ka)_{cut}=3.0 $. Similar results can be found, for example, in the
reference [\onlinecite{Tse}].

\section{Circular polarized light and temperature effect}

For circular polarized light the appropriate optical conductivity is

\end{subequations}
\begin{equation}
\sigma_{\pm }(\omega )\equiv\sigma _{xx}(\omega )\pm i\sigma _{xy}(\omega )
\end{equation}
where the longitudinal and charge Hall conductivity given by (\ref{Hall},\ref%
{xx}). Results for the absorptive part of the conductivity $Re\sigma _{\pm
}(\omega )$ are presented in Figure \ref{fig4} where we show separately $%
Re\sigma _{xx}$ (solid curve) for $s_{z}=+1$, and dotted curve for $s_{z}=-1$%
, to be compared with the dashed curve for $Im\sigma _{xy}$ with $s_{z}=+1$
and dash-dotted for $s_{z}=-1.$ For a given spin the onset in each pair of
curves for longitudinal and Hall conductivity are the same. The two begin to
deviate from each other as $\omega $ is increased above the threshold where
the Hall conductivity falls below its longitudinal value.

For the infinite band case we have seen in (\ref{Imxy}) that $Im\sigma _{xy}$
drops like $1/\omega $. Similar algebra for the infinite band limit gives
for the longitudinal conductivity

\begin{eqnarray}
&&Re\sigma _{xx}(\omega )=-\frac{e^{2}}{16\hbar }[1+\frac{(\Delta -\lambda
)^{2}}{\hbar ^{2}\omega ^{2}}]\theta (\hbar \omega -(\Delta -\lambda ))
\notag \\
&&\times \lbrack f(\frac{\lambda }{2}+\frac{\hbar \omega }{2}-\mu )-f(\frac{
\lambda }{2}-\frac{\hbar \omega }{2}-\mu )]
\end{eqnarray}

which drops less rapidly with increasing $\omega $ than does $Im\sigma
_{xy}(\omega )$ in equation (\ref{Imxy}) and provides a check in our
numerical results shown in Figure \ref{fig4} for a finite cut off $ka=3$.

The results in the top frame of Figure \ref{fig4} apply to temperature $%
T=10K $, while those in the bottom are for $T=300K$. For the specific value
of $\mu $ chosen i.e. $\mu =-0.845eV$ the chemical potential falls between
the two spin polarized valence bands. On comparing with the top frame we
note considerable temperature smearing of the spin up band. But the spin
down band by comparison is much less affected as it does not fall at the
Fermi energy but is everywhere below $\mu $ and hence does not respond to
temperature as effectively.

\begin{figure}[tp]
\begin{center}
\includegraphics[height=4.0in,width=3.0in]{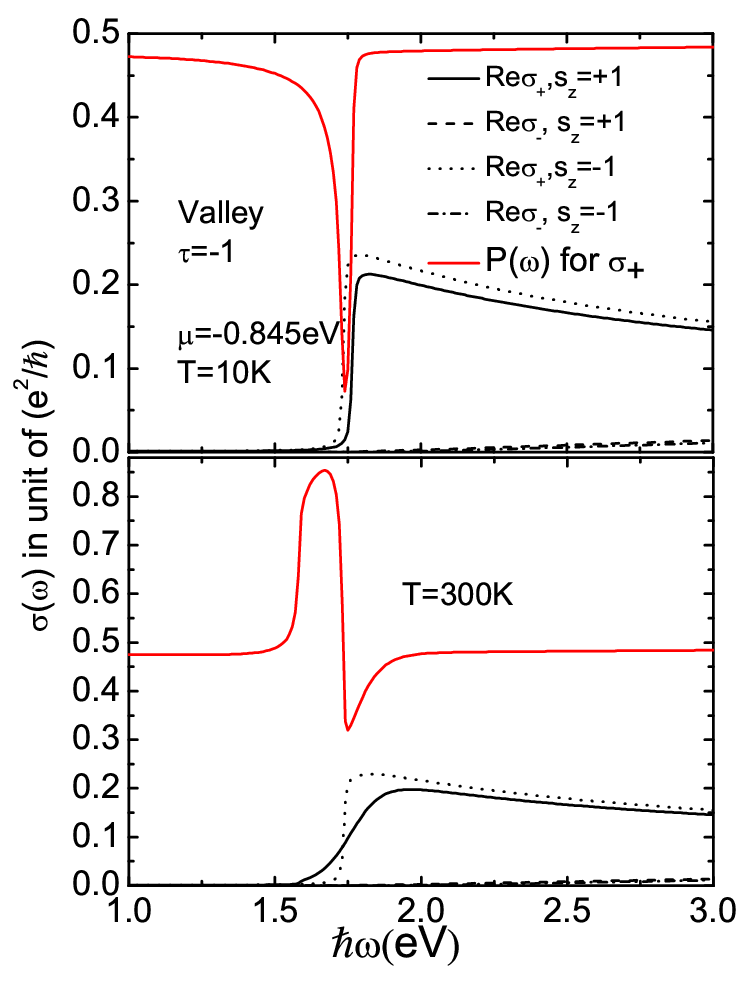}
\end{center}
\caption{Results for the absorptive part of the conductivity for circularly
polarized light, $\protect\sigma _{\pm }(\protect\omega)\equiv\protect\sigma %
_{xx}\pm i\protect\sigma _{xy}$. Here the valley index $\protect\tau =-1$.
The solid is for $Re\protect\sigma _{+}$ the dashed for $Re\protect\sigma %
_{-}$ both for spin index $s_{z}=+1$. For$s_{z}=-1$ the dotted curve is $%
\protect\sigma _{+}$ and the dashed dotted for $\protect\sigma _{-} $. The
dominant response in this valley is from $\protect\sigma _{+}$ right handed
circular polarization, also shown on the same figure is the up to up plus
down spin polarization for $\protect\sigma _{+}$ (solid red curve). The top
frame is at $T=10K$ and the bottom at $T=300K$. }
\label{fig5}
\end{figure}

In Figure \ref{fig5} we present our results for $Re\sigma _{\pm }(\omega )$
vs. $\omega $ in units of $e^{2}/\hbar $ for valley $\tau =-1$ in the case $%
\mu =-0.845eV$. The top frame applies to $T=10K$ while temperature is $300K$
in the bottom frame. Results are presented separately for $s_{z}=\pm 1$. We
note that $Re\sigma _{+}(\omega )$ is large and comparable in size for
either spin up or down while $Re\sigma _{-}(\omega )$ is very small in
comparison. This is understood from the optical selection rule which apply
to $\sigma _{xx}$ and $\sigma _{xy}$. In the infinite band limit we can show
that

\begin{eqnarray}
&&Re\sigma _{xx}(\omega )\mp Im\sigma _{xy}(\omega )=-\frac{e^{2}}{16\hbar }
\theta (\hbar \omega -(\Delta -\lambda ))\times  \notag \\
&&\lbrack 1\pm \frac{(\Delta -\lambda )}{\hbar \omega }]^{2}[f(\frac{\lambda
}{2}+\frac{\hbar \omega }{2}-\mu )-f(\frac{\lambda }{2}-\frac{\hbar \omega }{
2}-\mu )]  \label{sigma}
\end{eqnarray}
which implies a perfect cancelation at the onset energy $\omega =(\Delta
-\lambda )$. Here the equation (\ref{sigma}) is for $s_{z}=+1$ band with $%
-\lambda \rightarrow \lambda $ for $s_{z}=-1$ case. Effectively light
polarization provides valley selection to a very good approximation. It is
exact at the onset energy and remains quite good even at $\omega =4eV$, for
which energy the light polarization will reduce the absorption from this
valley to about $20\%$ its value for the opposite polarization. The
polarization selection rule found here for the dynamical conductivity agree
fully with those previous discussed by Yao et.al \cite{Yao} for inversion
symmetry breaking Hamiltonian in two valley systems, where they relate the
optical selection rule to the orbital magnetic moment and the Berry
curvature of particular valley considered. It is interesting to look at the
spin up spin down admixture of the optical response $Re\sigma _{+}(\omega )$
vs. $\omega $. We define
\begin{equation}
P(\omega )=\frac{\sigma _{+}^{spin\text{ }up}(\omega )}{\sigma _{+}^{spin
\text{ }up}(\omega )+\sigma _{+}^{spin\text{ }down}(\omega )}
\end{equation}
The results are shown as the red curve in Figure \ref{fig5}. We note that
for most frequencies $P(\omega )$ is close to $1/2$ but that around the
onset energy for $\sigma _{+}$ there is a region where $P(\omega )$ falls
below this value and can be close to $0$. This is easily understood as a
direct consequence of the displacement in onset between $Re\sigma
_{+}(\omega )$ for spin down (dotted curve) and for spin up (solid curve)
which implies a deficit of spin up electron in the region between these two
onsets. Temperature can have a large effect on the position and shape of $%
P(\omega )$ in the region of the onset as seen in the lower frame of Figure %
\ref{fig5}. This can be traced to the fact that the spin up electrons are
much more susceptible to temperature smearing for the case considered here
as we have already noticed. The solid curve ($\uparrow $) for $\sigma _{+}$
now extend to lower energies than does the short dashed curve ($\downarrow $%
). Temperature can in fact change the magnitude of $P(\omega )$ in the
region of interest from less than $1/2$ to larger than $1/2$ with this
entire region shifted toward lower energies. The peak in $P(\omega )$ for $%
T=300K$ is also broaden as compared with the valley in this same quantity
for $T=10K$. The spin admixture in this region can be manipulated with
temperature.

\section{Summary and Conclusions}

We presented expressions for the dynamic conductivity of $MoS_{2}$ based on
a simplified Hamiltonian which includes spin orbit coupling and band
structure parameters fit to first principle calculations. Using the Kubo
formula, the final expressions reduce to a sum over momentum $k$ centred
about the two valley points $K$ and $-K$ defining the corners of the
honeycomb lattice in the Brillouin zone. The bands are spin polarized and a
sum over spin and valley appears. The transverse conductivity is split into
charge, spin and valley Hall conductivity and, in all cases, depends on an
overlap of the Berry curvature multiplied with a common sum of two energy
denominators linear in frequency plus appropriate temperature factors and
channel dependent indices. The expression for the longitudinal conductivity
is not quite as simple but reduces to the common form around $k=0$ (zero
momentum). Simplified analytic expressions can be obtained in the infinite
band limit. We have presented numerical results for all these quantities as
a function of photon energy separating out, in each case, the contribution
from the separate spin channels. In the numerical calculations we use a cut
off on momentum of $ka=3.0$. The results are not qualitatively different
from the infinite band case although there are important quantitative
differences. The effect of temperature is considered. It provides a smearing
that tends to obscure the separate contributions to the conductivity. It
also can change very significantly the spin admixture in the frequency
window just above the main absorption threshold where there can be an
important imbalance between up and down spin, for a given valley when
circular polarized light is employed. The absorptive part of the
conductivity for right and left handed polarized light, which is the
appropriate quantity for valleytronics is also computed, and it is found
that second valley contributes very little to the absorption. For example it
is less than $20\%$ at $\omega =4eV$. with main absorption peak at $~1.7eV$.
The DC limit of the Hall conductivities is obtained analytically. For the
parameters appropriate to $MoS_{2}$, the real part of the valley Hall
conductivity is positive and large in value as compared with its spin Hall
counterpart which is negative for some values of the chemical potential. We
hope that our calculations can help further our understanding of the optical
properties of $MoS_{2}$.

\begin{acknowledgments}
This work was supported by the Natural Sciences and Engineering
Research Council of Canada (NSERC) and the Canadian Institute for
Advanced Research (CIfAR).
\end{acknowledgments}

\end{document}